\documentclass[twocolumn]{elsart}
\usepackage{graphicx,amsmath}
\begin{document}

\begin{frontmatter}
\title{Football goal distributions and extremal statistics}
\author{J.~Greenhough\corauthref{cor1}},
\ead{greenh@astro.warwick.ac.uk}
\author{P.~C.~Birch},
\corauth[cor1]{Corresponding author.}
\author{S.~C.~Chapman},
\author{G.~Rowlands}
\address{Department of Physics, University of Warwick, Coventry CV4 7AL, UK}

\begin{abstract}
We analyse the distributions of the number of goals scored by home teams, away teams, and the total scored in the match, in domestic football games from 169 countries between 1999 and 2001. The probability density functions (PDFs) of goals scored are too heavy-tailed to be fitted over their entire ranges by Poisson or negative binomial distributions which would be expected for uncorrelated processes. Log-normal distributions cannot include zero scores and here we find that the PDFs are consistent with those arising from extremal statistics. In addition, we show that it is sufficient to model English top division and FA Cup matches in the seasons of 1970/71 to 2000/01 on Poisson or negative binomial distributions, as reported in analyses of earlier seasons, and that these are not consistent with extremal statistics. 
\end{abstract}


\end{frontmatter}

\section{Introduction}\label{intro}
Few authors have considered football scores from a statistical point of view. Moroney \cite{moroney} showed that the numbers of goals scored by individual teams, and the total goal scores, were well described by a ``modification of the Poisson''; Reep et al. \cite{reep} later identify this as the negative binomial distribution, and found similar results for other ball games. Maher \cite{maher} then pointed out that a negative binomial distribution may arise from the aggregate of Poisson-distributed scores with a different mean for each team. The short-term predictability of results has subsequently been modelled using independent Poisson distributions with means dependent on teams' past performances \cite{dixon1,dyte}; an improved model \cite{dixon2} includes the scoring-rate dependence on both time and the current score.

Other aspects of the game have been examined including the effects of certain conditions on the scores - see \cite{norman} for a review.  Seeking a broader understanding, it has been suggested \cite{mala} that the distribution of goals per player may be linked with anomalous diffusion via the Zipf-Mandelbrot law. In this paper we show, in agreement with analyses of matches from the 1960s \cite{moroney,reep}, that 13,~000 English top division and 5,~000 FA Cup matches between the seasons of 1970/71 and 2000/01 \cite{facup} are closely-fitted by either Poisson or negative binomial distributions. However, we find that the number of goals scored by home and away teams, and the total goals, in over 135,~000 domestic football games (leagues and cups, hereafter referred to as domestic matches) from 169 countries between 1999 and 2001 \cite{rsssf}, cannot be fitted over their entire ranges by Poisson or negative binomial distributions. Instead, we find that the data can be modelled by extremal statistics (explained in Sect.~\ref{stats}).

The ubiquity of power-law relationships in both nature \cite{bak} and the field of econophysics \cite{mantegna,stanley1,stanley2,solomon} has spawned a significant amount of literature in recent years. Intriguingly, extremal statistics in a global measure are found in turbulent fluids and other highly-correlated systems \cite{bram1,bram2,aji,bram3,zheng,bram4}. Hence the significance and origin of extremal and power-law-tailed distributions are currently of considerable interest in statistical physics; the use of probability distributions in the modelling of complex systems is a topical approach to the inverse problem. From an operational perspective, knowledge of the statistics would be an important constraint on any model for the game.

\section{The probability density function (PDF)}\label{PDF}
The first step in our analysis of each data set is the construction of its PDF. The PDF $P(x)$ of a variable $X$ is defined such that the probability that $X$ lies within a small interval $dx$ centred on $X=x$, is equal to $P(x)dx$. $P(x)$ is normalised so that
\begin{equation}
\int_{x_{min}}^{x_{max}}P(x)dx=1.\label{eq1}
\end{equation}
Here, $x$ takes the integer values of goals scored ($x_{min}=0$ and $x_{max}$ is the maximum number of goals scored in the sample of matches) so Eq.~\ref{eq1} becomes $\sum_{x_{min}}^{x_{max}}P(x)=1$. We further normalise each PDF to the sample mean $\mu$ and standard deviation $\sigma$ to enable comparison with extremal distributions (see Sect.~\ref{stats}).

The Poisson distribution is defined by
\begin{equation}
P(x\mid\mu)=\frac{\mu^{x}}{x!}e^{-\mu}I_{(0,1,...)}(x), \label{poisseq}
\end{equation}
where $I_{(0,1,...)}(x)$ ensures that $P(x)=0$ for non-integer $x$. In the Poisson PDF $\mu =\sigma^2$; for data to be well-fitted by this distribution we require $\mu\approx\sigma^2$. It is explained in \cite{moroney,reep} that this condition does not hold for football goals because a constant probability per unit time of scoring a goal is not a valid assumption. Instead, a \emph{compound Poisson} or \emph{negative binomial} distribution is used, defined by
\begin{equation}
P(x\mid r,p)=\binom{r+x-1}{x}p^{r}q^{x}I_{(0,1,...)}(x) \label{nbineq}
\end{equation}  
where $x$ is the number of goals scored with probability $q$ per goal before $r$ ``failed goals'' (probability $p=1-q$) have occured. The negative binomial PDF has $\mu =r(1-p)/p$ and $\sigma^{2}=r(1-p)/p^{2}$; fitting to data thus requires $\mu /\sigma^{2}=p\leq 1$ and $\mu p/(1-p)=r$ where we round $r$ to the nearest integer.

\section{Extremal statistics}
\label{stats}
Our data analysis presented in Sect.~\ref{res} (Figs.~1 to 3) shows that the tails of the PDFs of goal scores in the domestic matches clearly deviate from both the Poisson and the negative binomial distributions. Here, we compare the PDFs of the data with those arising from extremal statistics. We choose extremal distributions fitted over the entire dataset in preference to a piecewise fit of arbitrary functions as (1)~they have been observed in a wide variety of natural systems; (2)~they may suggest a physical interpretation, as they arise in situations where only the largest events are observed; (3)~following normalisation of the data, only one parameter ($a$) remains to be estimated, and (4)~unlike log-normal PDFs, they can be applied to data including zero values.

The two limiting distributions of interest are ``Gumbel's asymptote'' and Fr\'{e}chet \cite{fisher,gum,sorn}. In outline, the limiting distributions result from selecting the maximum value $x_{max}$ from each of a large number of large samples whose individual members are drawn from a distribution $P(x)$. When $P(x)$ decreases more rapidly than any power-law (as $x\rightarrow\infty$), ``Gumbel's asymptote'' has the form
\begin{equation}
P_{G}(x_{max})=K(e^{u-e^{u}})^{a}
\end{equation}\label{gumeq}
$\mathrm{with}\hspace{5mm}u=b(x-s)$

where in the limit of an infinite number of measurements $a\equiv 1$; the constants $K$, $b$, and $s$ are fixed by normalisation as in Sect.~\ref{PDF} (see \cite{bury,chapman}). Selecting the \emph{second} largest values from the same large samples produces a PDF of the same functional form as Eq.~\ref{gumeq} but with $a\equiv 2$. 

Fr\'{e}chet distributions $P_{F}(x_{max})$ arise in the same manner when the underlying PDF $P(x)$ is power-law; the power-law tail of this underlying distribution is preserved in the Fr\'{e}chet, thus lending itself to the fitting of heavy-tailed data. Mathematically, $P_{F}(x_{max})$ can be  defined by Eq.~\ref{gumeq} but with $u=\alpha+\beta\ln(1+x/G)$, where $K$, $\alpha$, and $G$ are again fixed by normalisation, and $\beta=(1-a)^{-1}$ \cite{chapman}. These curves exist for $a>1$.

A simple heavy-tailed distribution often encountered in nature is the log-normal. Log-normal distributions with the same means and variances as the datasets provide very poor models in all cases if scores of zero are included. The domestic home and away scores are quite well modelled by log-normal PDFs \emph{providing zero scores are neglected}. Alternatively, one goal can be added to all scores but, since the log-normal is not invariant under translation, the results are no more meaningful. Scores of zero occur frequently and should not be removed; we seek a single heavy-tailed PDF appropriate for modelling integer data from zero upwards.

\section{Results}\label{res}
\begin{figure*}
\centering
\includegraphics[width=\textwidth]{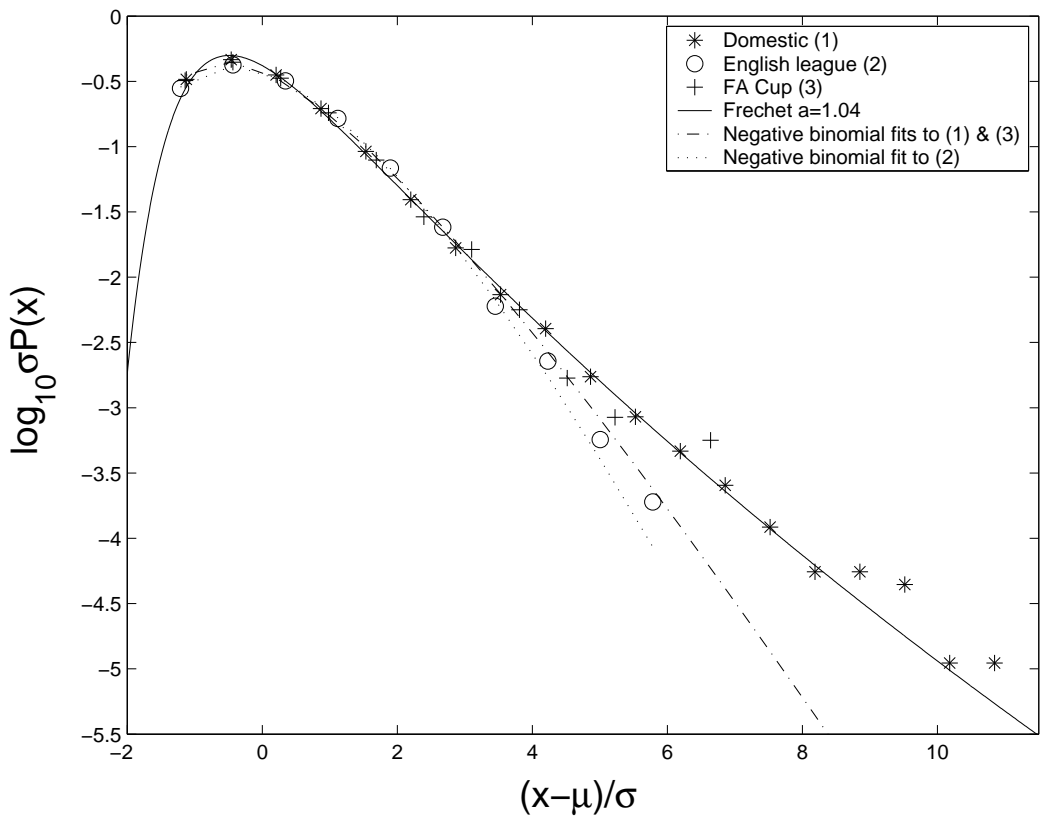}
\caption{PDFs of goals scored by home teams, normalised with respect to $\mu$ and $\sigma$, showing how domestic matches are more closely fitted by a Fr\'{e}chet distribution than by a negative binomial. Coincident curves are plotted as a single line as indicated in the legend.}\label{home}
\end{figure*}

\begin{figure*}
\centering
\includegraphics[width=\textwidth]{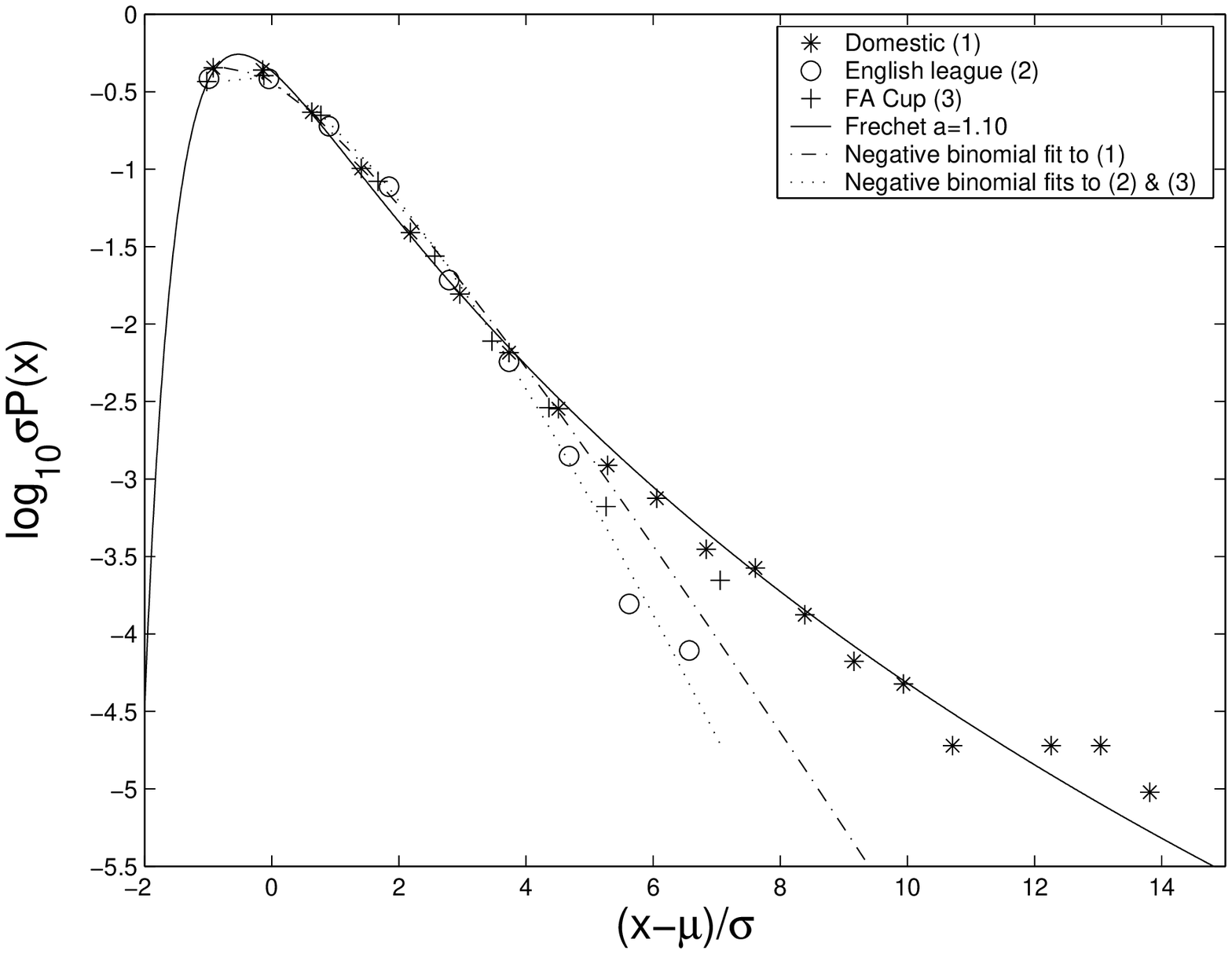}
\caption{PDFs of goals scored by away teams, normalised with respect to $\mu$ and $\sigma$, showing how domestic matches are more closely fitted by a Fr\'{e}chet distribution than by a negative binomial. Coincident curves are plotted as a single line as indicated in the legend.}\label{away}
\end{figure*}

\begin{figure*}
\centering
\includegraphics[width=\textwidth]{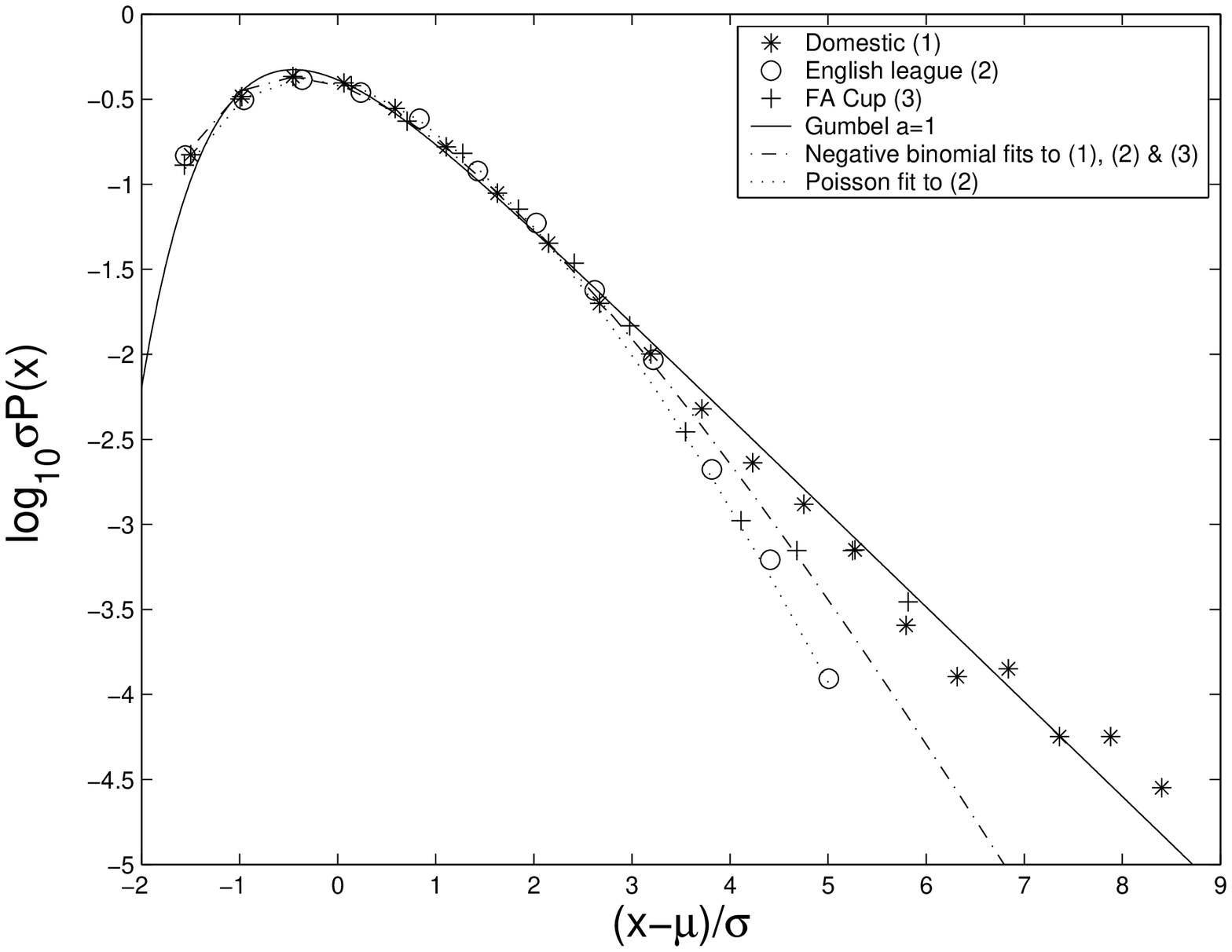}
\caption{PDFs of total goals scored, normalised with respect to $\mu$ and $\sigma$, showing how domestic matches are more closely fitted by a Gumbel  distribution than by a negative binomial. Coincident curves are plotted as a single line as indicated in the legend. Domestic $\mu=2.9$, $\sigma=1.9$; league $\mu=2.6$, $\sigma=1.7$; Cup $\mu=2.8$, $\sigma=1.8$.}\label{total}
\end{figure*}

As discussed in Sect.~\ref{intro}, the Poisson distribution has been demonstrated to be inferior to the negative binomial when modelling football scores; only where this is \emph{not} the case do we include a Poisson fit in Figs.~\ref{home}-\ref{total}. In Fig.~\ref{home} we show the PDFs of home team scores with their respective negative binomial PDFs (fitted to $\mu$ and $\sigma$) along with the best-fit extremal distribution (see Sect.~\ref{stats}) for the domestic matches.
While the league scores follow a negative binomial PDF, it is clear that the domestic scores are better described by a Fr\'{e}chet distribution beyond about $\mu +3\sigma$ (a home score of about 6 goals). Although the Cup scores are suggestive of some departure from a negative binomial PDF, we cannot quantify the functional form of this tail. Counting errors caused by binning a finite dataset are omitted from Figs.~\ref{home}-\ref{total} for the purpose of clarity; typical sizes of counting errors are indicated by fluctuations around a smooth trend and become apparent in the final few bins.

We plot the away team scores in Fig.~\ref{away}. 
Again, the domestic scores are consistent with a Fr\'{e}chet distribution above $\mu +4\sigma$ (an away score of about 6 goals) whereas negative binomial PDFs suffice for the league and Cup scores if the last few points are discounted as explained above. The total goal scores with fitted negative binomial PDFs are plotted in Fig.~\ref{total}. Here we find that the domestic scores are consistent with a Gumbel distribution (see Sect.~\ref{stats}) above $\mu +3\sigma$ (9 goals), and that the league scores are more suggestive of a Poisson than a negative binomial PDF.   

We now provide more detailed analysis of the goodness of fit of the extremal PDFs to the domestic scores. Figures~4-6 show in linear form the closeness of fit of various distributions to the domestic data. To quantify whether the data are consistent with the fitted PDFs, one must estimate the likely counting errors in the numbers of points in the bins (introduced by the finite size of the dataset). We are interested in the distribution of the number of points in a bin, given both the total number of data points and the probability that any point will lie in that particular bin (given directly by the fitted PDF). A binomial distribution of counting errors is a reasonable estimate, and one can thus estimate the upper and lower limits of the number of points one would expect to find in any bin on $95\%$ of occasions; these are plotted with dashed lines. The ending of a dashed line indicates that the lower/upper limit of the expected number of occurences of the corresponding score is zero; where both limits stop, no higher scores are expected given the size of the dataset. From these plots we again conclude that a Fr\'{e}chet a=1.04 PDF is the best fit to domestic home scores, a Fr\'{e}chet a=1.10 PDF is the best fit to domestic away scores, and a Gumbel a=1 PDF is the best fit to domestic total scores.

\begin{figure*}		
\centering
\includegraphics[width=\textwidth]{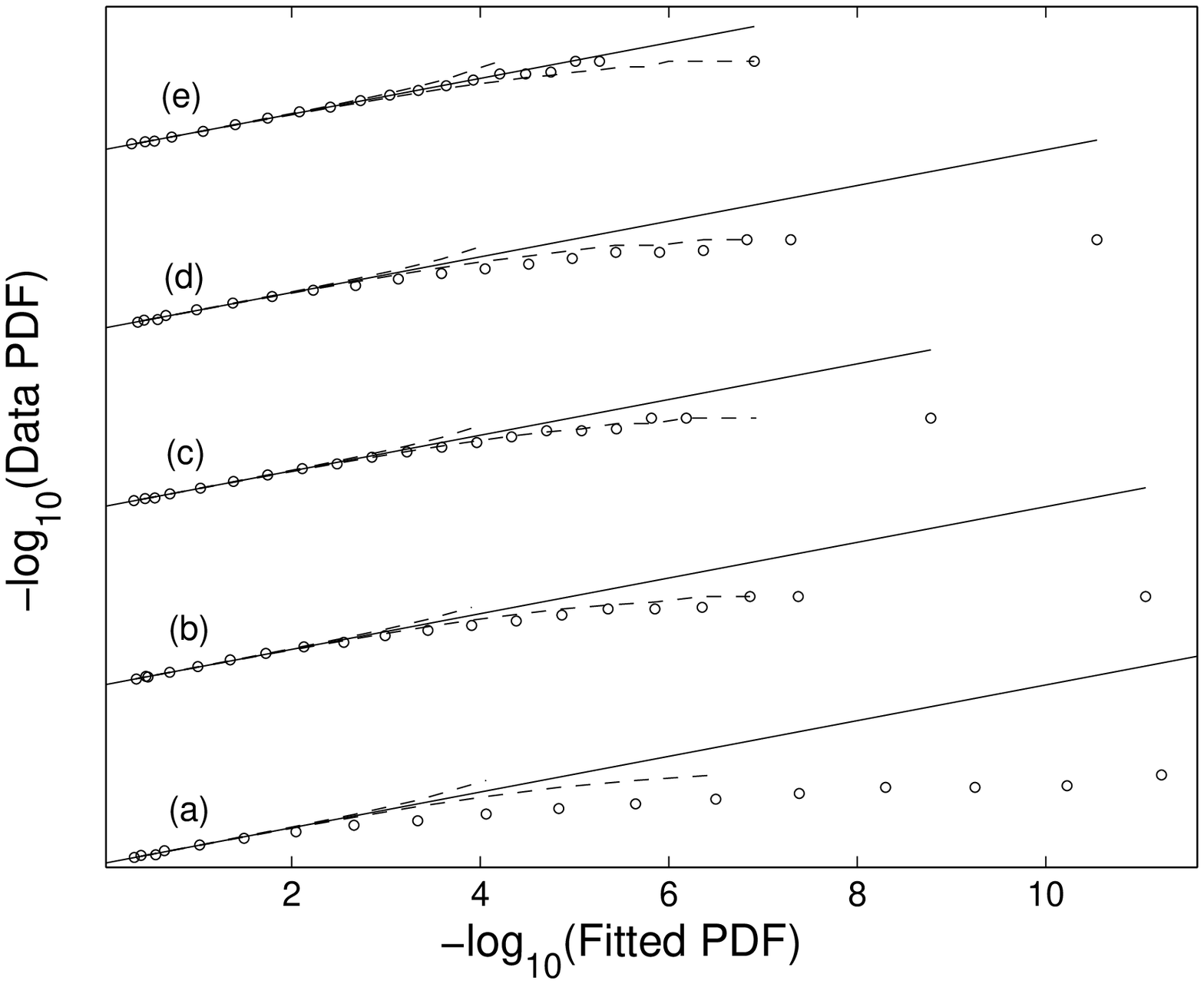}
	\caption{Normalised PDF of domestic home scores plotted against a range of fitted PDFs. Straight lines indicate where the points would lie were the fits perfect, and are separated by an arbitrary vertical displacement; dashed lines indicate 95\% binomial counting errors. Note how the Fr\'{e}chet a=1.04 PDF (e) provides a superior fit to the gumbel a=1,2 (c,d), negative binomial (b), and Poisson (a) distributions; compare Fig.~\ref{home}.}
\end{figure*}

\begin{figure*}
\centering
\includegraphics[width=\textwidth]{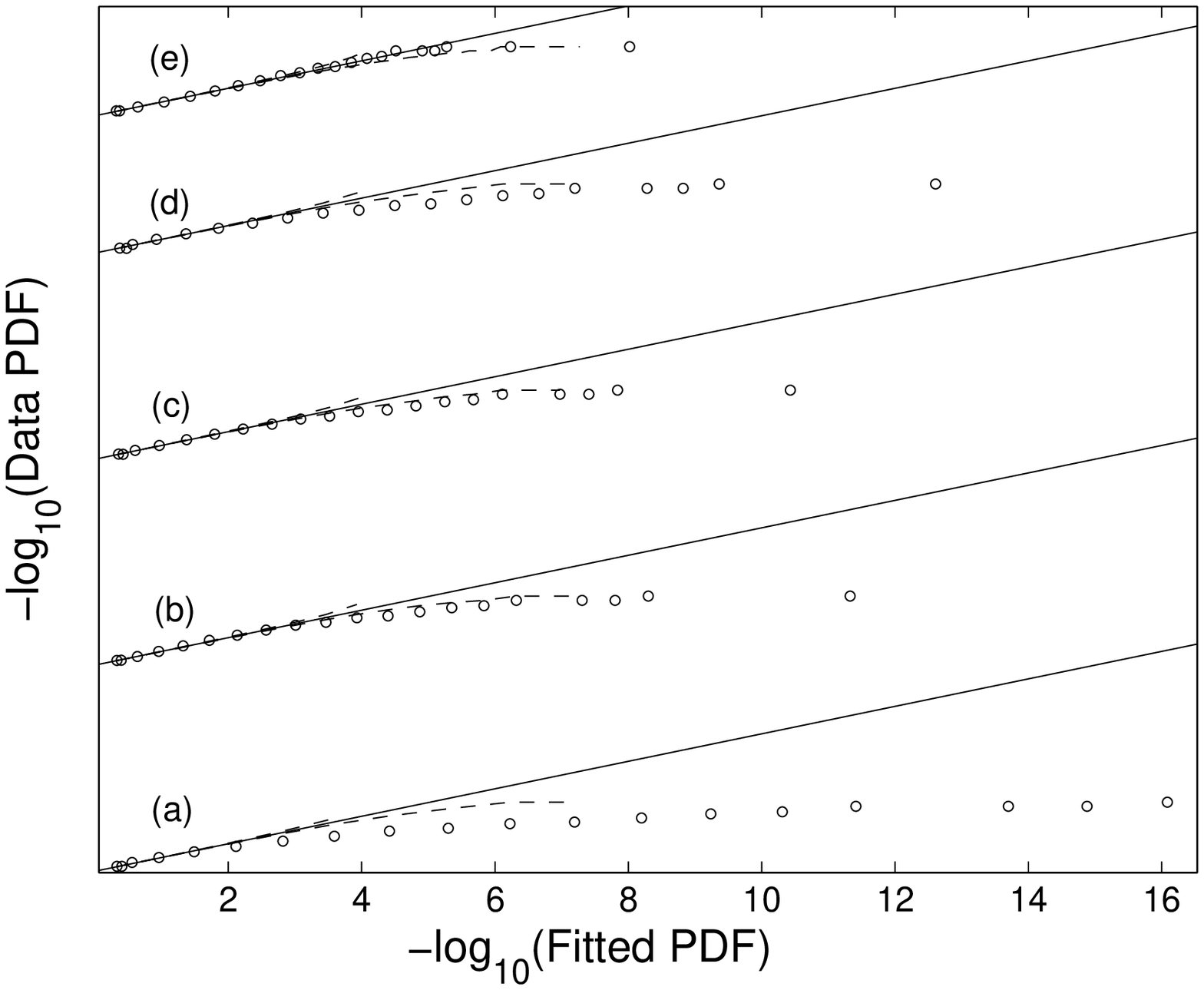}
	\caption{Normalised PDF of domestic away scores plotted against a range of fitted PDFs. Straight lines indicate where the points would lie were the fits perfect, and are separated by an arbitrary vertical displacement; dashed lines indicate 95\% binomial counting errors. Note how the Fr\'{e}chet a=1.10 PDF (e) provides a superior fit to the gumbel a=1,2 (c,d), negative binomial (b), and Poisson (a) distributions; compare Fig.~\ref{away}.}
\end{figure*}

\begin{figure*}		
\centering
\includegraphics[width=\textwidth]{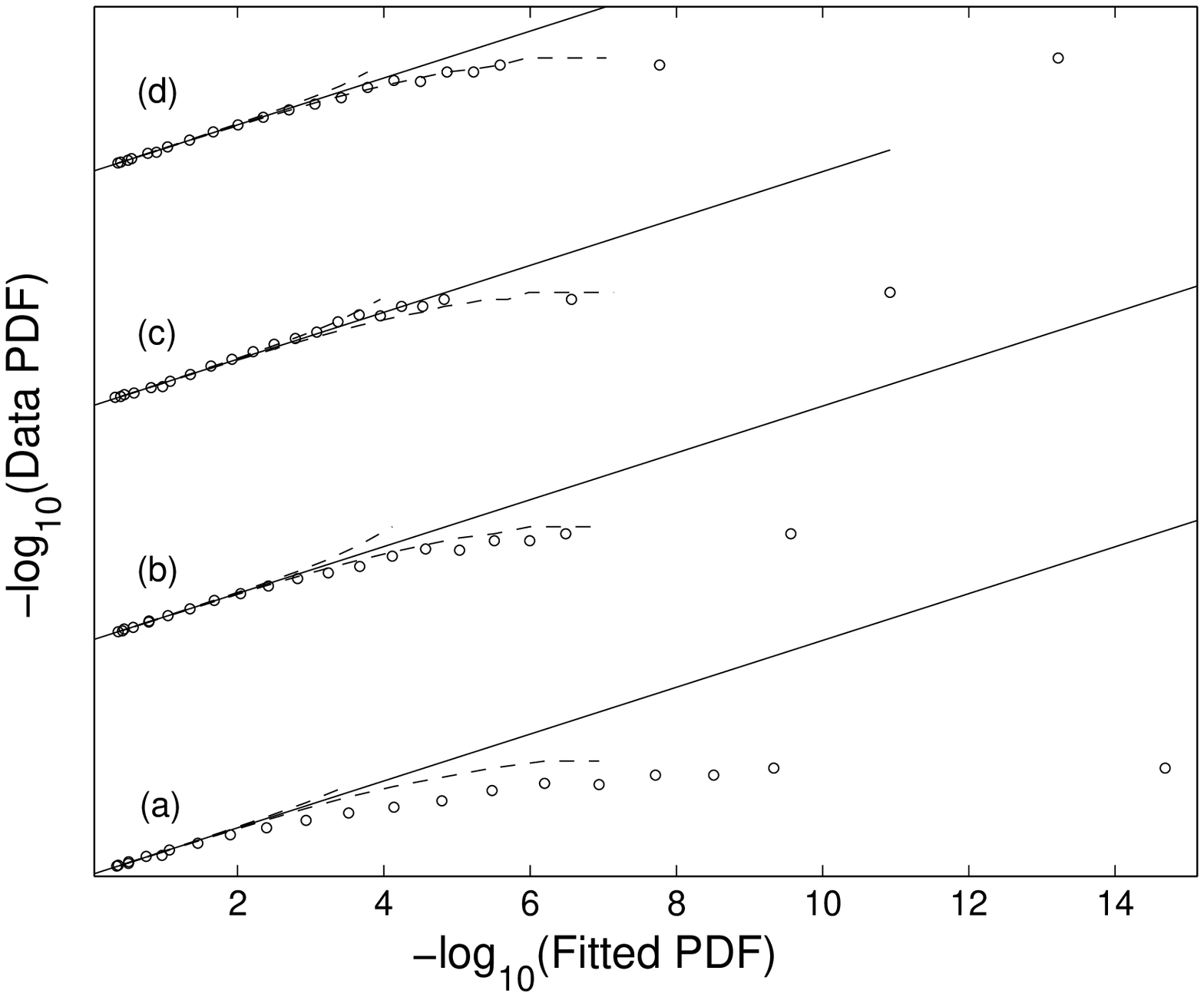}
	\caption{Normalised PDF of domestic total scores against a range of fitted PDFs. Straight lines indicate where the points would lie were the fits perfect, and are separated by an arbitrary vertical displacement; dashed lines indicate 95\% binomial counting errors. Note how the Gumbel a=1 PDF (c) provides a superior fit to the gumbel a=2 (d), negative binomial (b), and Poisson (a) distributions; compare Fig.~\ref{total}.}
\end{figure*}

The empirical PDF varies between the negative binomial and extremal PDF in each case; for low scores both the negative binomial and extremal distributions provide satisfactory fits. However, as shown in the previous figures, there is a strong departure from the negative binomial on to heavier-tailed distributions for the higher scores. Our aim here is to identify a single distribution that fits the whole dataset rather than an arbitrary piecewise fit. The latter could always be achieved given a sufficient number of independent distribution functions to fit to different ranges of data, but would ultimately be less informative of the underlying processes. 

In this context it is important to note that the distribution of the aggregate of many thin-tailed datasets (i.e. the pooled data) is heavy-tailed if the variances of the component datasets differ \cite{granger,kon}. Hence the heavy-tails seen in worldwide football results could arise simply from the aggregation of scores from many teams. Individual teams' scores may follow different Poisson distributions, which when pooled produce countries' scores following negative binomials, and then the aggregation of countries' scores is heavy-tailed. Testing this hypothesis would require significantly more data than used here, and would run over an interval of time that may imply significant changes in the game process. The alternative -- and more interesting -- possibility is that the heavy tails are the result of some inherent process that increases the likelihood of high scores over their Poisson-based expectations. 
 
We also find that both the English data and the worldwide domestic results show a mean goal difference (home score minus away score) of 0.51, an aggregate home advantage (see \cite{clarke}), and a bias towards uneven scores as the total score rises is evident in the larger domestic dataset; these trends are well-known in the world of football.

It is important to note that the observation of a departure from negative binomial distributions is not the result of a larger dataset for domestic matches. Whilst more rare events are observed in a larger sample and the distribution extends to higher values with lower probabilities, it is nevertheless possible to distinguish between the different distributions, as we have shown,  without considering these extreme values. We have looked briefly at other individual countries and find similar trends to those shown for English matches.

\section{Conclusions}
We have shown that the simplest models -- the thin-tailed Poisson and negative binomial distributions based on the assumption of uncorrelated processes -- do not fit domestic (worldwide) football matches between 1999 and 2001 beyond the low scores. Heavier-tailed distributions are required if these datasets are to be fitted with single PDFs. Log-normal distributions do not include zero scores whereas extremal distributions can model the entire range of scores. Extremal distributions have been observed in a variety of complex systems and our results may then inform the modelling of football games.

In addition, using English top division and FA Cup matches in the seasons of 1970/71 to 2000/01, we confirm the Poisson or negative binomial nature of English scores as reported in analyses of earlier football seasons.  

\begin{ack}
We are grateful to Nick Watkins and Richard Dendy for helpful discussions. JG and PCB acknowledge Research Studentships from the UK Particle Physics and Astronomy Research Council. Data provided by members of the Recreational Sport Soccer Statistics Foundation and SportingData.com. 
\end{ack}

\end{document}